\documentclass[showpacs,twocolumn,prl,superscriptaddress]{revtex4-1}

\usepackage{slashed}
\usepackage{mathrsfs}
\usepackage{amsmath}
\usepackage{amssymb}
\usepackage{revsymb}
\usepackage{graphicx,accents}
\usepackage{graphicx,epstopdf}
\usepackage{mathrsfs}
\usepackage{bm}
\usepackage{psfrag}
\usepackage{hyperref}
\hypersetup{colorlinks=true,citecolor=Red,urlcolor=Red}
\usepackage{bbm}
\usepackage{bbold}
\usepackage{tabularx}
\usepackage{graphicx}
\usepackage[normalem]{ulem}
\usepackage[usenames,dvipsnames]{xcolor}

\usepackage[usenames,dvipsnames]{xcolor}
\usepackage{multirow,tabularx}
\usepackage{xcolor,pifont}
\newcommand*\colourcheck[1]{%
  \expandafter\newcommand\csname #1check\endcsname{\textcolor{#1}{\ding{52}}}%
}

\newcommand{\be}{\begin{equation}}
\newcommand{\ee}{\end{equation}}
\newcommand{\bea}{\begin{eqnarray}}
\newcommand{\eea}{\end{eqnarray}}

\newcommand{\trm}[1]{\textrm{#1}}

\newcommand{\vphi}{\varphi}

\newcommand{\ud}{\mathrm{d}}
\newcommand{\vtheta}{\vartheta}

\definecolor{bk1}{RGB}{0,200,100}

\begin{document}


\title{Plasma harmonic generation for highly efficient Breit-Wheeler pair creation}

\author{Suo Tang}
\email{tangsuo@ouc.edu.cn}
\affiliation{College of Physics and Optoelectronic Engineering, Ocean University of China, Qingdao, Shandong, 266100, China}


\date{\today}
\begin{abstract}
Observation of efficient Breit-Wheeler electron-positron pair creation is one of the main goals for modern-day strong laser-particle experiments.
We propose exploring this process by colliding plasma-generated harmonics, from currently available laser pulses, with a beam of GeV photons.
We show that the creation yield is enhanced for more than one order of magnitude, compared with the yield obtained without high-harmonic generation.
The robustness of the yield enhancement is demonstrated by considering multiple interaction parameters and the potential photon source for the upcoming experiments.
Moreover, we show that due to the evident field asymmetry in the plasma-generated harmonics, the spin-polarization of the created pairs could be as high as $60\%$.
\end{abstract}
\maketitle

Breit-Wheeler (BW) pair creation---the decay of light quanta into an electron-positron pair, 
is one of the most intriguing strong-field quantum electrodynamic (QED) phenomena~\cite{NBW1934,ritus85,RMP2012Piazza,Fedotov:2022ely,RMP2022_045001} and one of the main goals for modern-day strong laser-particle experiments~\cite{Joshi_2018,Jacobs:2021fbg,Borysova2021,Abramowicz:2021zja,naranjo2021pair,chen22,Macleod_2022,Turner:2022hch}.
The current high-power laser facilities in operation or in development have entered the multipetawatt regime and pave the way to measure the process in the nonlinear and nonperturbative regime~\cite{Sung:17,Kiriyama:18,Shen_2018,Danson2019PetawattAE,Yoon:21}.
This process has been investigated theoretically in different type of laser fields~\cite{nikishov64,PRA2013_052125,2016PRD085028,PRD013010,Titov:2018bgy,ilderton2019coherent,Ilderton:2019vot,PRD076017,SeiptPRA052805,NJP_2021,PRD2021_016009,TangPRA2021,Tang:2021qht,Tang:2022a,TangPRD056003,BenPRD116015} to improve the creation yield, tailor the spectra of the created particles, and 
{recently suggested to generate polarized positron beams in ultraintense finely-tuned asymmetric fields~\cite{PRADaniel061402,PRLchen174801,WAN2020135120}.}
A robust scheme to generate brilliant, highly polarized positron beams is crucial for testing the physics beyond the standard model~\cite{MOORTGATPICK2008131,ILC1}.

The typical scenario for observing the BW pair creation is to collide a beam of high-energy probe photons with a strong laser field.
In the center-of-mass frame of the collision, there is an energy threshold $2m$ must be satisfied, 
where $m$ is the electron's rest mass and the natural units $\hbar=c=1$ is used throughout.
To overcome this threshold, one can improve either the energy of the probe photon,
or the intensity of the laser pulse to provide more laser photons for the creation.
The yield of the process thus depends mainly on two factors: i) the parameter $\eta=k\cdot \ell/m^2$ characterizing the center-of-mass energy of the collision and the number of laser photons $n\sim 2/\eta$ needed for the creation; ii) the classical intensity \mbox{$\xi_{l}=e\mathcal{E}_{l}/m\omega_{l}$} of the laser pulse characterizing the nonlinearity of the creation,
where $\ell^{\mu}$ is the photon momentum, $k^{\mu}$, $\mathcal{E}_{l}$ and $\omega_{l}$ are respectively the laser's wavevector, intensity and carrier frequency of the laser, and $e$ is the positron charge.

In the upcoming laser-particle experiments, \emph{e.g.} LUXE~\cite{Abramowicz:2021zja} and FACET II-E320~\cite{chen22}, high-energy photons with $E_{\gamma}\sim10~\trm{GeV}$ and laser pulses with the intensity $\xi_{l}\sim O(1)$ and carrier frequency $\omega_{l}\sim 1~\trm{eV}$ would be utilized.
The BW yield of the potential measurement in these campaigns would be limited as $\eta\ll 1$ and $\eta\xi \sim 1$.
To improve the particle yield, we propose to collide the high-energy photons with the plasma-generated harmonics in the reflection from the currently available laser pulse interacting with an overdense plasma~\cite{PoP871619,PRE046404,RMP.81.445,Tang_2019}.
The energy parameter is considerably improved by the rich high-order harmonics, and due to the locking phase of the low-order harmonics~\cite{2004PRL115002,2011PRE046403,Bulanov_2013}, the pulse amplitude is effectively magnified.
The creation yield could thus be significantly enhanced compared with that in the initial pulse without the harmonic generation.
Moreover, the created particles could be highly polarized by the asymmetric harmonic field when proper oblique incidence of the laser pulse is considered.

\begin{figure}[t!!!]
 \center{\includegraphics[width=0.43\textwidth]{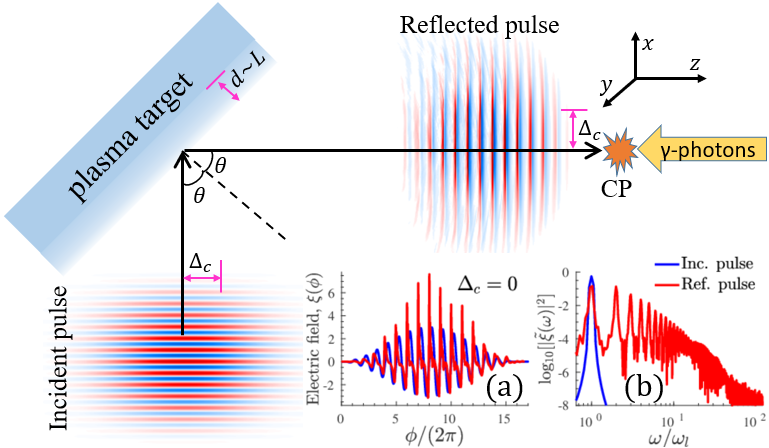}}
\caption{Sketch for efficient Breit-Wheeler pair creation using plasma generated harmonics: a strong laser pulse irradiates an overdense plasma with the incident angle $\theta$ and gets reflected to collide with a beam of high-energy probe photons at the collision point (CP). The contour plots are the electric field of the incident pulse with $\xi_{l}=3$ and reflected pulse resulting from a 2D PIC simulation for the incidence $\theta=45^{\circ}$. In the insets, the comparisons between the incident and reflected pulse at the pulse center \mbox{$\Delta_{c}=0$} are given: (a) the electric fields and (b) the corresponding frequency spectra.
The detail of the simulation is given in the main text.
}
\label{Fig_Sketch}
\end{figure}
The sketch of our proposal is shown in Fig.~\ref{Fig_Sketch}: a linearly polarized laser pulse, irradiating an overdense plasma with the incident angle $\theta$, gets reflected at the plasma surface and then collides nearly head-on with a beam of high-energy photons at the collision point.
The periodically oscillating current driven by the incident pulse at the plasma surface, modulates the incident field and emits rich high-order harmonics into the reflected field~\cite{RMP.81.445}.
The electric fields of the incident and reflected pulse are compared in Fig.~\ref{Fig_Sketch} as an example for the incident angle $\theta=45^{\circ}$, resulting from a 2D particle-in-cell (PIC) simulation using the code EPOCH~\cite{Arber_2015}:
The incident laser pulse, \mbox{$\xi(\phi,\Delta_{c})=\xi_{l}\sin(\phi)\sin^{2}\left[\phi/(2N)\right]\exp(-\Delta^{2}_{c}/\sigma^{2})$}, is in the duration $0<\phi<2N\pi$, where $\xi_{l}=3$, $N=16$, $\phi=k\cdot x$ is the laser phase of the carrier frequency, $\sigma=6\lambda_{l}$ corresponds to the full-width-at-half-maximum about $10\lambda_{l}$ and the laser wavelength is $\lambda_{l}=0.8\mu\trm{m}$;
The plasma target has an overdense bulk $n_{e}=15n_{c}$ and a pre-gradient $n_{c}\exp(d/L)/2$, where $L=\lambda_{l}/11$, $d$ denotes the distance to the plasma surface, $0<d<L\ln(2n_{e}/n_{c})$, $n_{c}=m\omega^{2}_{l}/(4\pi e^2)$ is the plasma critical density. 
As shown, the amplitude of the reflected pulse is effectively magnified, see Fig.~\ref{Fig_Sketch}~inset (a) for the electric field at the pulse center $\Delta_{c}=0$, and rich high-order harmonics of the incident frequency are superposed in the reflected pulse, see Fig.~\ref{Fig_Sketch}~inset (b).
As we will show later, this effective amplitude magnification and rich harmonic generation could result in orders of magnitude enhancement of the particle yield.
The other main feature of the reflected pulse as shown in Fig.~\ref{Fig_Sketch} inset (a) is the evident field asymmetry, which would then lead to high spin polarization for the created particles.

The full QED calculation for the BW process has been done and outlined as follows:
the positron spectrum produced by a high-energy photon in the scaled potential \mbox{$a(\phi) =-\int^{\phi}_{0} \xi(\phi')\ud\phi'$} can be written as
\begin{align}
\frac{\ud \trm{P}}{\ud s}&=\frac{i\alpha}{2\pi \eta}  \int \ud\vphi \int\frac{\ud\vtheta}{\vtheta}~e^{\frac{i\vtheta\Lambda}{2\eta s(1-s)}}\left( 1-\vtheta^2\langle \xi \rangle^{2}h_{s}\right)\,,\label{Eq_prob_polar0}
\end{align}
where $\alpha=e^2\approx 1/137$ is the fine structure constant, \mbox{$s=k\cdot q/ k\cdot \ell$} is the fraction of the lightfront momentum taken by the positron, $h_{s}=(2s^{2}-2s+1)/[4s(1-s)]$, $\vphi$ ($\vtheta$) is the average (interference) phase and $\Lambda=1-\langle a\rangle^{2} + \langle a^2\rangle$ is the Kibble mass~\cite{kibble64,dinu16,king19a,AntonPRD085002}.
The window average $\langle f\rangle$ is calculated as $\langle f\rangle=\frac{1}{\vtheta}\int^{\vphi+\vtheta/2}_{\vphi-\vtheta/2}f(\phi)\ud \phi$.
In the collision geometry, the plasma reflected pulse propagates in the $z$-direction with the magnetic (electric) field in the $y$ ($x$)-direction.
The spin of the created positrons is polarized in the direction of the magnetic field as
\begin{align}
 S_{y}(s) &=\frac{1}{\ud\trm{P}/\ud s} \frac{\alpha}{2\pi\eta} \int\ud\vphi \int\ud\vtheta~e^{\frac{i\vtheta\Lambda}{2\eta(1-s)s}} \frac{ \langle \xi \rangle}{2 s}\,,
\label{Eq_spin}
\end{align}
with the total polarization $S_{y}$ acquired by integrating Eq.~(\ref{Eq_spin}) without $\ud\trm{P}/\ud s$ but normalized with the total yield $\trm{P}$.
From Eq.~(\ref{Eq_spin}), we can infer that the spin of the created positrons in the pulse with symmetric field must be unpolarized as $S_{y} \propto \langle \xi \rangle \approx 0$~\cite{PRA023417,Del_Sorbo_2018}.
The detailed derivation and calculation of Eqs.~(\ref{Eq_prob_polar0}) and~(\ref{Eq_spin}) can be found in Ref.~\cite{Tang:2022a}.
Here, the plane-wave approximation is used as the scale of the field transverse variation is much larger than the electron Compton wavelength and the particles' transverse displacement in the near head-on collision~\cite{DiPiazza2015PRA,DiPiazza2016PRL,DiPiazza2017PRA032121,DiPiazza2021PRD076011}.

\begin{figure}[t!!!]
 \center{\includegraphics[width=0.48\textwidth]{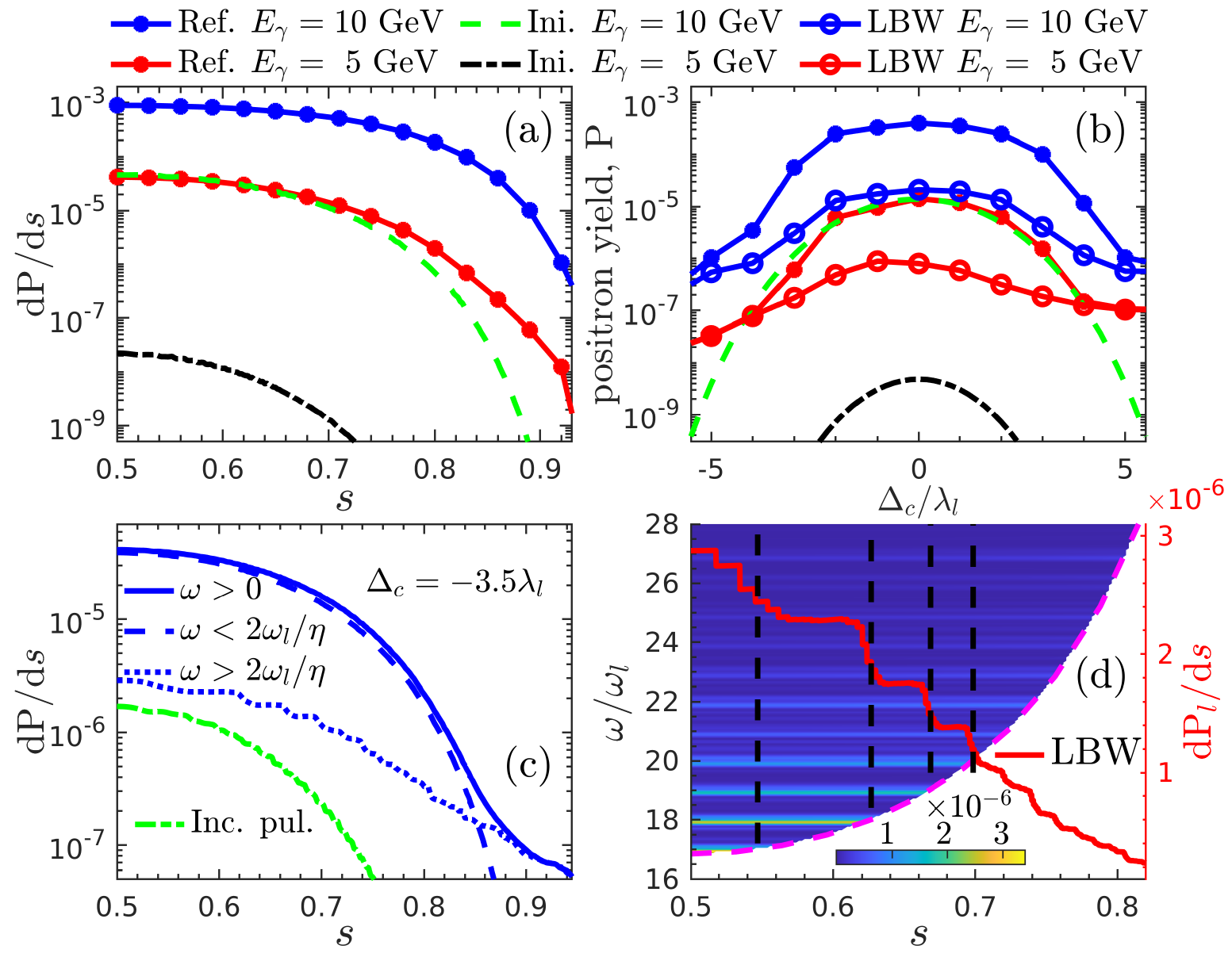}}
\caption{Breit-Wheeler pair creation in the plasma reflected pulse. (a) Positron spectra created by the probe photons ($E_{\gamma}=5, 10~\trm{GeV}$) colliding head-on with the incident and reflected pulses at the center $\Delta_{c}=0$.
 (b) Positron yield \emph{vs} transverse misalignment $\Delta_{c}$ in the incident and reflected pulses. The yield of the linear Breit-Wheeler (LBW) process in the reflected pulse is also given.
 (c) Contributions of the different frequency components in the reflected pulse to the total spectrum created by a $10~\trm{GeV}$ photon at $\Delta_{c}=-3.5\lambda_{l}$. The spectrum in the initial pulse is plotted for comparison.
 (d) Double differential spectrum $\ud^{2} \trm{P}_{l}/(\ud s \ud \omega)$ and energy spectrum $\ud \trm{P}_{l}/\ud s $ of the LBW pair creation. The magenta dashed line is its frequency threshold, $\omega_{*}=\omega_{l}/[2\eta s(1-s)]$, and the vertical black dashed lines denote the boundary of the contribution from different order of harmonics $\omega/\omega_{l}=17,18,19,20$ for $E_{\gamma}=10~\trm{GeV}$.
 The collision between the plasma harmonics and probe photons is $30\lambda_{l}$ away from the plasma surface.
 }
\label{Fig_yield}
\end{figure}
In Fig.~\ref{Fig_yield} (a), we compare the positron spectra created by the high-energy photons (\mbox{$E_{\gamma}=5, 10~\trm{GeV}$}, corresponding to \mbox{$\eta=0.06,~0.12$} respectively) in the head-on collision with the incident and reflected pulses at $\Delta_{c}=0$: the spectrum created in the reflected pulse is much higher and decays much slower than that in the incident pulse.
This means significant yield enhancement with the plasma reflected pulse as shown in \mbox{Fig.~\ref{Fig_yield} (b)}, in which the positron yield with the change of the transverse misalignment $\Delta_{c}$ is given.
For the $10~\trm{GeV}$ photon, the enhancement at $\Delta_{c}=0$ is about $30$ times, and for $E_{\gamma}=5~\trm{GeV}$, the yield is enhanced for more than three orders of magnitude and even slightly larger than that created by the $10~\trm{GeV}$ photon in the incident pulse.

This significant yield enhancement can be attributed to the rich harmonics contained in the reflected pulse.
In \mbox{Fig.~\ref{Fig_yield} (c)}, we show the positron spectra created by the different frequency components in the reflected pulse by filtering out the other frequency components artificially.
For the relatively lower-order harmonics $\omega/\omega_{l}<2/\eta$, which come from the Doppler frequency upshift during the backward acceleration of the plasma surface~\cite{RMP.81.445}, the creation is triggered by absorbing more than one harmonic photons, $n\sim 2\omega_{l}/(\eta \omega)>1$, \emph{i.e.} nonlinear BW process.
The spectrum of this nonlinear process, shown as the blue dashed line in \mbox{Fig.~\ref{Fig_yield} (c)}, is much higher than the total spectrum (green dotted-dashed line) obtained in the incident pulse, due to two factors: i) the considerable decrease of the photon number absorbed for the creation, and ii) the effective increase of the reflected field amplitude superposed by these phase-locking low-order harmonics~\cite{PRL095004}.
For the higher-order harmonics $\omega/\omega_{l}>2/\eta$, which come from the synchrotron radiation of the well-accelerated surface layer~\cite{PRL245005} and improve the collision energy parameter considerably, $\eta\to \omega\eta/\omega_{l}>2$,
the creation can thus be triggered with only one harmonic photon, \emph{i.e.} linear BW process, which is forbidden in the incident pulse with the normal frequency.
The spectrum of this linear process [blue dotted line in \mbox{Fig.~\ref{Fig_yield} (c)}] could also be much higher than the total spectrum obtained in the incident pulse.

We point out that the total spectrum [blue solid line in \mbox{Fig.~\ref{Fig_yield} (c)}] acquired in the reflected pulse can be divided numerically into the contributions from the linear and nonlinear processes, \emph{i.e.} $\trm{P}\approx\trm{P}_{\omega<2\omega_{l}/\eta} + \trm{P}_{\omega>2\omega_{l}/\eta}$ as the intensity of high-order harmonics ($\omega/\omega_{l}>2/\eta$) is too weak to affect the nonlinear process and the low-order harmonics ($\omega/\omega_{l}<2/\eta$) are excluded from the linear process by the frequency threshold [see Eq.~(\ref{Fig_pert})].
The nonlinear process is more important in the energy regime $s\to 0.5$ and determines the height of the total spectrum, while the linear process is more important in the regime $s\to 1$ (and $s\to 0$) resulting in the slow spectral decay at the high (low) energy tail.
Around the pulse center $\Delta_{c}\to 0$ where the laser intensity is high, the pair creation is dominated by the nonlinear BW process as shown in \mbox{Fig.~\ref{Fig_yield} (b)}.
The linear process is dominant in the region away from the pulse center, where the multiphoton absorption is strongly suppressed by the lower intensity, and thus results in the effective pair creation in a much broader transverse region than that in the incident pulse, also shown in \mbox{Fig.~\ref{Fig_yield} (b)}.
For lower-energy photons, the importance of the linear BW process is higher as the nonlinear process is limited with the absorption of more harmonic photons.

In \mbox{Fig.~\ref{Fig_yield} (c)}, a clear step-structure is observed in the spectrum of the linear BW process.
This structure results directly from the harmonic structure in the frequency spectrum of the reflected pulse, and can be clearly seen via Eq.~(\ref{Eq_prob_polar0}) after doing the perturbative expansion \mbox{($\xi\to 0$)} and keeping only $\xi^{2}$ terms as:
\begin{align}
\frac{\ud \trm{P}_{l}}{\ud s}&=\frac{\alpha\omega_{l}}{\pi\eta}\int^{\infty}_{\omega_{*}} \frac{\ud\omega}{\omega^{2}} |\tilde{\xi}(\omega)|^2\left(h_{s}+\omega_{*}\frac{\omega-\omega_{*}}{\omega^{2}}\right)\,,
\label{Fig_pert}
\end{align}
which gives the positron spectrum [red solid line in \mbox{Fig.~\ref{Fig_yield} (d)}] matching exactly the full QED calculation for $\omega/\omega_{l}>2/\eta$ in \mbox{Fig.~\ref{Fig_yield} (c)}, where $\tilde{\xi}(\omega)=\int\ud\phi\xi(\phi)e^{i\phi\omega/\omega_{l}}$, and $\omega_{\ast} = \omega_{l}/[2\eta s(1-s)]$ is the threshold frequency required to create a pair [magenta dashed line in \mbox{Fig.~\ref{Fig_yield} (d)}].
From Eq.~(\ref{Fig_pert}), we can see that the harmonic peaks could bring the horizontal fringes into the double differential spectrum, \mbox{$\ud^{2} \trm{P}_{l}/(\ud s \ud \omega) \propto |\tilde{\xi}(\omega)|^2$}, shown in \mbox{Fig.~\ref{Fig_yield} (d)}.
After integrating over the pulse frequency, a clear step-structure is imprinted into the positron spectrum as lower-order harmonics cannot contribute at larger $s$, see the boundaries (vertical black dashed lines) of the contribution from different order of harmonics in \mbox{Fig.~\ref{Fig_yield} (d)}.
This harmonic structure may provide the idea to observe the linear BW process by colliding the high-energy photon with the plasma reflected pulse after filtering out its low-frequency components. (To filter out low-frequency components, one can refer to~\cite{Tsakiris_2006,Horlein_2010}.)

In Fig.~\ref{Fig_yield}, the horizontal axis of the spectra are restricted in the region $0.5 < s < 1$ due to the symmetry, $\trm{P}(s) = \trm{P}(1-s)$ in Eq.~(\ref{Eq_prob_polar0}).
The slight asymmetry at $\Delta_{c}=0$ in Fig.~\ref{Fig_yield} (b) for the yield in the reflected pulse is because of the mild deformation of the plasma surface.
The convergence of the harmonics due to plasma denting surface~\cite{Vincenti_2014}, is not critical as the steep plasma pre-gradient and intermediately intense laser pulse with broad transverse profile are applied. 
The collision point between the reflected pulse and photon beam is $30\lambda_{l}$ away from the plasma reflection point, and no notable difference is observed when the collision moves further.


\begin{figure}[t!!!]
 \center{\includegraphics[width=0.44\textwidth]{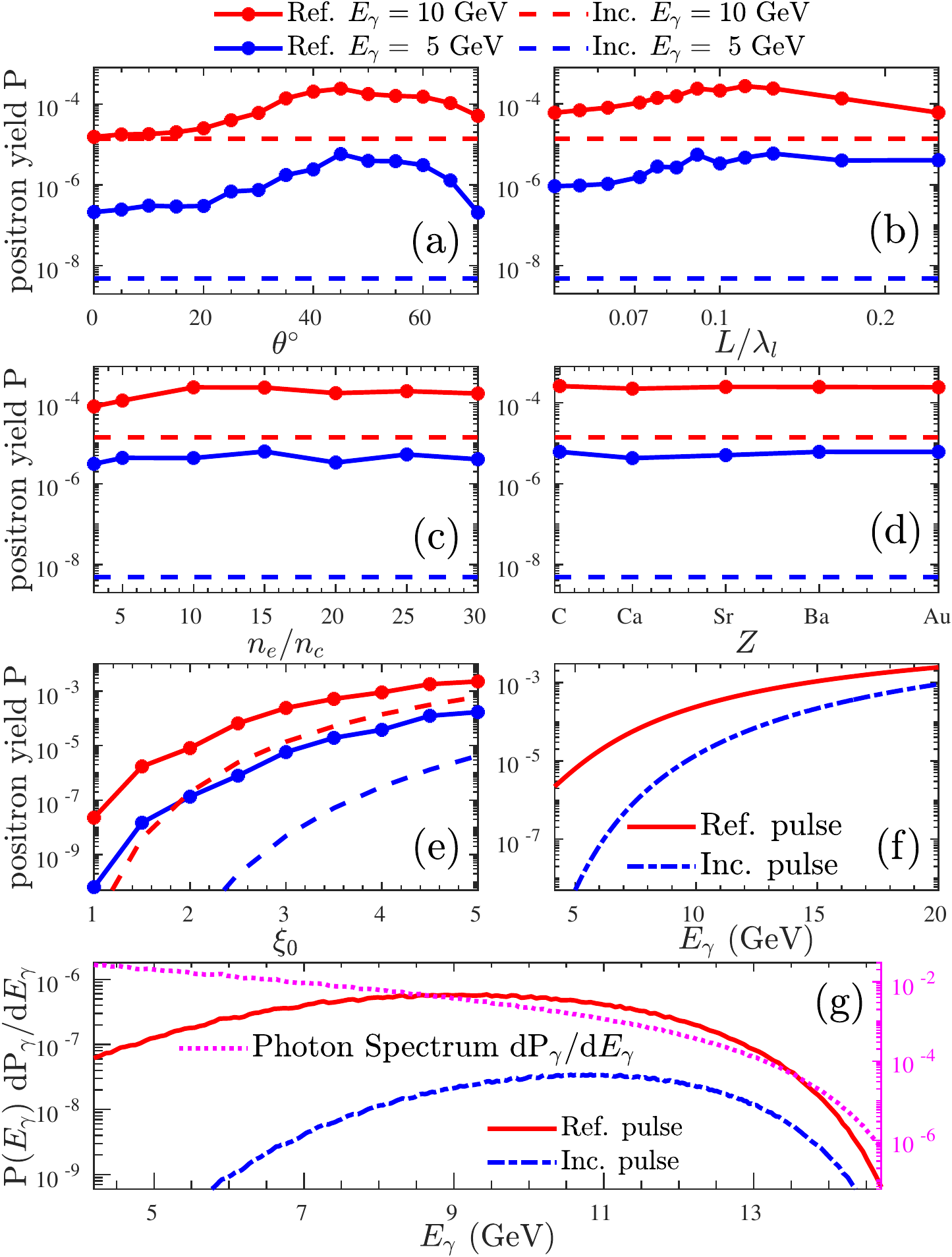}}
\caption{Robustness of the yield enhancement with the plasma generated harmonics by considering multiple interaction parameters: (a) laser incident angle $\theta$; (b) plasma pre-gradient $L$; (c) plasma bulk density $n_{e}$; (d) plasma material. Fully ionized materials: C, Ca, Sr, Ba, Au with respectively the charge number $Z=6,20,38,56,79$, are used;
(e) laser intensity $\xi_{l}$; (f) photon energy $E_{\gamma}$ and (g) Compton photon source scattered by a $16.5~\trm{GeV}$ electron colliding head-on with the incident pulse. 1D PIC simulations are performed for harmonic generation.
If not specified in each panels, the following parameters: $\xi_{l}=3$, $\theta=45^{\circ}$, $L=\lambda_{l}/8$, $n_{e}=15n_{c}$ and $Z=79$ are used. In (e), the plasma bulk density increases with the laser intensity as $n_{e}=5\xi_{l}n_{c}$.
The corresponding results in the incident pulse without harmonics are given for comparison.
}
\label{Fig_robustness}
\end{figure}

The robustness of the yield enhancement with the plasma reflected pulse is tested in Fig.~\ref{Fig_robustness} for multiple interaction parameters.
The plasma harmonic generation is simulated with series of 1D PIC setups~\cite{Obl_Bourdier} for weakly focused pulses $\sigma\to \infty$. 
The positron yield in the plasma reflected pulse is significantly enhanced for a broad range of incident angle $\theta$ and plasma pre-gradient $L$ as shown in Figs.~\ref{Fig_robustness} (a) and (b) with the peak enhancement at around $\theta=45^{\circ}$ and $L=\lambda_{l}/10$, at which the efficiency of the harmonic generation is maximized~\cite{Thaury_2010,Shikha_2023}.
For the highly overdense plasma $n_{e}\gg n_{c}$ with a proper pre-gradient, its bulk density and target material may not affect the efficiency of harmonic generation considerably, but only affect the peak widening of high-order harmonics via the hole-boring effect~\cite{TangPRE2017}.
The yield enhancement is thus nearly stable for the high bulk density in~\mbox{Fig.~\ref{Fig_robustness} (c)} and different target materials in \mbox{Fig.~\ref{Fig_robustness} (d)}.
For the relatively lower density $n_{e}\sim \xi_{l}n_{c}$ in \mbox{Fig.~\ref{Fig_robustness} (c)}, the yield enhancement becomes smaller as the harmonic generation is reduced with the lower plasma current.
In \mbox{Figs.~\ref{Fig_robustness} (e) and (f)}, the level of the yield enhancement with the plasma harmonics is declined for the larger laser intensity and higher photon energy.
This may be because i) the decreasing of the photon number needed for the creation due to the harmonics becomes less important, since more laser photons are provided with the larger laser intensity and less laser photons are needed for the higher photon energy,
and ii) the importance of the linear BW contribution from the high-order harmonics becomes weaker.
However, in the intermediate intensity region $\xi_{l}\sim O(1)$ with the moderate photon energy $E_{\gamma}\lesssim 10~\trm{GeV}$, the yield enhancement with the plasma harmonics is still significant as shown in \mbox{Fig.~\ref{Fig_robustness} (e)}: at $\xi_{l}=5$, the yield is enhanced by about $5$ times for $E_{\gamma}=10~\trm{GeV}$ and about $40$ times for $E_{\gamma}=5~\trm{GeV}$.
This implies that the plasma-generated harmonics have the great advantage for the current laser-particle experiments, in which the dominant energy of the potential photon sources is much smaller than $10~\trm{GeV}$~\cite{Abramowicz:2021zja,naranjo2021pair,chen22,MacLeodPRA2023}.
In \mbox{Fig.~\ref{Fig_robustness} (g)}, the spectrum, $\ud \trm{P}_{\gamma}/\ud E_{\gamma}$ of the Compton photon source scattered by a $16.5~\trm{GeV}$ electron colliding with the incident pulse ($\xi_{l}=3$) is considered~\cite{TangPRA022809,TANG2020135701,PRD096004}. As shown, with the plasma reflected pulse, the energy spectrum, $\trm{P}(E_\gamma)~\ud \trm{P}_{\gamma}/\ud E_{\gamma}$ of the created positrons is significantly improved with the total yield enhanced for more than $20$ times.
The final yield could be further improved if the laser pulse for the Compton scattering is also replaced with the plasma-generated harmonics.

\begin{figure}[t!!!]
 \center{\includegraphics[width=0.45\textwidth]{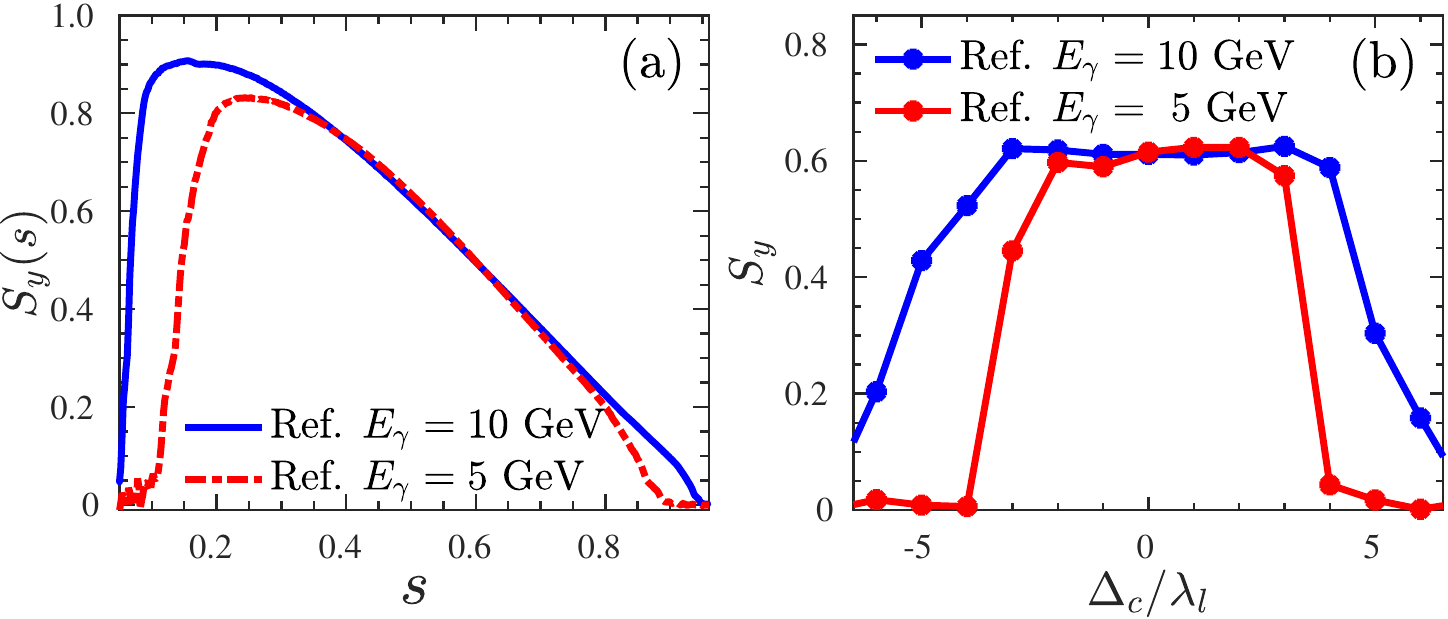}}
\caption{Spin polarization of the created positrons by the high-energy photons with \mbox{$E_{\gamma}=5,~10~\trm{GeV}$} in the plasma reflected pulse with the incident angle $\theta=45^{\circ}$.
(a) Spectrum of the positron's spin polarization at $\Delta_{c}=0$.
(b) Variation of the total spin polarization with the change of the misalignment $\Delta_{c}$.
 The same parameters are used as in Fig.~\ref{Fig_yield}.}
\label{Fig_spin}
\end{figure}

The other advantage with the plasma-generated harmonics is to create the positron beam with the high spin polarization.
This is obtained due to the evident field asymmetry as $S_{y} \propto \langle \xi \rangle$ in Eq.~(\ref{Eq_spin}), when the laser oblique incidence is applied for the harmonic generation, see Fig.~\ref{Fig_Sketch} inset (a) for the incident angle $\theta=45^{\circ}$.
The generation of the harmonics in the oblique incidence is pronounced in one of the half laser cycle in which the oscillation of the surface current is promoted by the laser electric field, and suppressed in the other half cycle as the current oscillation is weakened by the electric field~\cite{RMP.81.445}.
As shown in Fig.~\ref{Fig_spin} for $\theta=45^{\circ}$, the spin of the created positrons is highly polarized in the $+y$ direction, in which the magnetic field is magnified [The electric field is magnified in the $+x$ direction, see Fig.~\ref{Fig_Sketch} inset (a)].
In \mbox{Fig.~\ref{Fig_spin} (a)}, the spin polarization of the created positron with the peak higher than $80\%$ around $s=0.2$, decreases with the increase of the energy as $S_{y}(s) \propto s^{-1}$ in Eq.~(\ref{Eq_spin}), and at the low (high) energy limit $s\to 0$ ($s\to 1$), the spin polarization goes to zero because of the dominance of the linear BW contribution.
This can be seen via the perturbative approximation, $S_{y}(s)\propto \xi^{3} \to0$ at $\xi\to 0$.
As the pairs are more probable to be created around $s=0.5$, the total polarization of the created positrons around the pulse center is about $62\%$ shown in Fig.~\ref{Fig_spin} (b). 
In the region away from the pulse center, the spin polarization decays to zero quickly.
As the asymmetry of the harmonic field depends on the laser incident angle, the positrons' spin polarization may be simply controlled by adjusting the laser incidence.

In conclusion, we detailed the exploration of the Breit-Wheeler pair creation using the plasma generated harmonics.
The rich harmonics can not only improve the yield of the nonlinear Breit-Wheeler process significantly, but also open an effective channel for the linear Breit-Wheeler process.
In the currently available region of laser intensity and photon energy, the particle yield could be improved for more than one order of magnitude. This proposal could be realized at upcoming experiments and used as a robust scheme to generate brilliant, highly polarized positron beams.
We point out that the efforts, to improve the harmonic-generation efficiency~\cite{PRL125001,yeung2017experimental,PRL264803,PRL244801}, can also be devoted here to further enhance the particle yield.

\begin{acknowledgments}
The author thanks Dr. A. Ilderton and Dr. B. King for useful discussion and careful reading of the manuscript.
The author acknowledges the support from the Natural Science Foundation of China, Grant No.12104428.
The work was carried out at Marine Big Data Center of Institute for Advanced Ocean Study of Ocean University of China.
\end{acknowledgments}

\bibliographystyle{apsrev}
\providecommand{\noopsort}[1]{}

\end{document}